\documentclass[fleqn,usenatbib]{mnras}
\usepackage{newtxtext,newtxmath}
\usepackage[T1]{fontenc}
\usepackage{ae,aecompl}
\usepackage{here}
\usepackage{float}

\usepackage[dvipdfmx]{graphicx}	% Including figure files
\usepackage{amsmath}	% Advanced maths commands
\usepackage{lscape}
\usepackage{graphicx}
\usepackage{multirow}
\usepackage{color}
\usepackage{bm}
\usepackage{comment}
\usepackage{ulem}

\newcommand{\Msun}{\mathrm{M}_{\odot}} 
\newcommand{\Zsun}{\rm{Z}_{\odot}}
\newcommand{\MJup}{\mathrm{M}_{\rm J}}

\newcommand{\Secref}[1]{Section\:\ref{#1}}

\newcommand{\Tabref}[1]{Table\:\ref{#1}}
\newcommand{\Eqref}[1]{equation\:(\ref{#1})}

\newcommand{\Figref}[1]{Fig.\:\ref{#1}}

\newcommand{\Panelsref}[4]{Figs.\:\ref{#1}#2 and \ref{#3}#4}

\title[Planet formation in a subsolar-metallicity disc]{Formation of a wide-orbit giant planet in a gravitationally unstable subsolar-metallicity protoplanetary disc}
\author[R. Matsukoba et al.]
{Ryoki Matsukoba,$^{1}$\thanks{E-mail: r.matsukoba@tap.scphys.kyoto-u.ac.jp} 
Eduard I. Vorobyov,$^{2,3}$
Takashi Hosokawa$^{1}$, and Manuel Guedel$^2$
\\
$^{1}$Department of Physics, Graduate School of Science, Kyoto University, Sakyo, Kyoto 606-8502, Japan\\
$^{2}$Department of Astrophysics, University of Vienna, T\"{u}rkenschanzstrasse 17, 1180, Vienna, Austria\\
$^{3}$Research Institute of Physics, Southern Federal University, Roston-on-Don 344090, Russia
}

\date{Accepted XXX. Received YYY; in original form ZZZ}

\pubyear{2023}

\begin{document}
\label{firstpage}
\pagerange{\pageref{firstpage}--\pageref{lastpage}}
\maketitle

%%% word limit: 250 %%%
\begin{abstract}
Direct imaging observations of planets revealed that wide-orbit ($>10$\:au) giant planets exist even around subsolar-metallicity host stars and do not require metal-rich environments for their formation. A possible formation mechanism of wide-orbit giant planets in subsolar-metallicity environments is the gravitational fragmentation of massive protoplanetary discs. Here, we follow the long-term evolution of the disc for 1\:Myr after its formation, which is comparable to disc lifetime, by way of a two-dimensional thin-disc hydrodynamic simulation with the metallicity of 0.1\:$\Zsun$. We find a giant protoplanet that survives until the end of the simulation. The protoplanet is formed by the merger of two gaseous clumps at $\sim$0.5\:Myr after disc formation, and then it orbits $\sim$200\:au from the host star for $\sim$0.5\:Myr. The protoplanet's mass is $\sim$10\:$\MJup$ at birth and gradually decreases to 1\:$\MJup$ due to the tidal effect from the host star. The result provides the minimum mass of 1\:$\MJup$ for protoplanets formed by gravitational instability in a subsolar-metallicity disc. We anticipate that the mass of a protoplanet experiencing reduced mass loss thanks to the protoplanetary contraction in higher resolution simulations can increase to $\sim$10\:$\MJup$. We argue that the disc gravitational fragmentation would be a promising pathway to form wide-orbit giant planets with masses of $\ge1$\:$\MJup$ in subsolar-metallicity environments.
\end{abstract}
\begin{keywords}
planets and satellites: formation -- protoplanetary discs -- hydrodynamics -- methods: numerical
\end{keywords}

%%%%%%%%%%%%%%%%%%%%%%%%%%%%%%%%%%%%%%%%%%%%%%%%%%
%%%%%%%%%%%%%%%%% BODY OF PAPER %%%%%%%%%%%%%%%%%%

%%%%%%%%%%%%%%%%%%%%%%%%%%%%%%%%%%%%%%%%%%%%
%%%%%%%%%%%%%%%%%%%%%%%%%%%%%%%%%%%%%%%%%%%%
%%% SECTION 1 %%%%
\section{Introduction}
\label{Sec:1}

%---------------------------------------------------%

Recent advancements in observations have led to the discovery of over 5000 exoplanets \citep[the NASA Exoplanet Archive;][]{Akeson2013}. In particular, observations of giant planets have progressed in the past decade by combining multiple observation methods, such as radial velocity, transit, microlensing, and direct imaging, revealing the presence of giant planets in a variety of environments with orbital radii ranging from $\sim10^{-2}$ to $10^3$\:au (e.g. \citealt{Zhu2021} for a review).

%---------------------------------------------------%

In addition to the increase in the number of exoplanet discoveries, observations of central stars with a planetary system have also advanced, and statistical studies have been conducted to examine the correlation between the characteristics of planets and their host stars. An interesting result of those studies is the relationship between the mass of giant planets and the metallicity of their host stars. For example, gas giant planets increase in mass (and radius) as the metallicity of their host star increases \citep{Buchhave2014, Santos2017, Petigura2018, Narang2018, Schlaufman2018}. However, this trend is reversed when the planet mass exceeds $\sim$4\:$\MJup$, and the metallicity of the host star begins to decreases as the planet mass increases \citep{Santos2017, Narang2018, Schlaufman2018, Maldonado2019}. Furthermore, such a negative correlation is stronger in wide-orbit ($>10$\:au) giant planets discovered through direct imaging, with the turning point located around 5\:$\MJup$ \citep{Swastik2021}. Thus, the difference in metallicity dependence between Jupiter-mass ($<4$--5\:$\MJup$) and super-Jupiter planets ($>4$--5\:$\MJup$) suggests that they have different formation mechanisms.

%---------------------------------------------------%

There are two major mechanisms for the formation of giant planets, one of which is the core accretion mechanism. In this mechanism, a rocky planetary core is first formed and then grows into a giant planet through the rapid gas accretion \citep{Mizuno1980, Bodenheimer1986}. The minimum core mass required for the onset of the accretion depends on several factors, such as radial distance from the host star, core accretion rate, grain opacity and so on \citep[e.g.][]{Pollack1996, Ikoma2000, Rafikov2006}, and declines with radial distance, from $\sim8$\:${\rm M}_\oplus$ at 5\:au to $\sim5$\:${\rm M}_\oplus$ at 100\:au \citep{Piso2014, Piso2015}. Since the core accretion mechanism requires a large mass planetary core, it is preferable in a solid-rich and high metallicity environment. Therefore, the core accretion mechanism is favourable for the formation of Jupiter-mass planets, but not for super-Jupiter planets. 

%---------------------------------------------------%

The other mechanism for the formation of giant planets is the gravitational instability. This mechanism forms giant planets through the fragmentation of a massive self-gravitating protoplanetary disc \citep{Boss1998, Rice2003, Mayer2007, Kratter2010, Machida2011, Zhu2012, Vorobyov2013, Tsukamoto2015, Nayakshin2017, Stamatellos2018, Vorobyov2018}. Disc fragmentation is considered to occur with a similar frequency in low-metallicity environments (\citealt{Tanaka2014, Matsukoba2022}, see also \citealt{Bate2014, Bate2019}), particularly in the metallicity range of $-0.8<[{\rm Fe/H}]<0.6$ that is characteristic for observed giant planets \citep{Adibekyan2019}, and is more suitable for forming subsolar-metallicity giant planets than the core accretion mechanism. Additionally, disc fragmentation tends to occur at a disc scale of $\sim100$\:au \citep[e.g.][]{Rafikov2005, Boley2009, Rice2010, Takahashi2016, Jin2020}, which is advantageous for the formation of wide-orbit giant planets.

%---------------------------------------------------%

Numerical calculations and analytical studies considered the formation of giant planets through gravitational instability in subsolar-metallicity protoplanetary discs \citep{Boss2002, Matsuo2007, Meru2010, Vorobyov2020-9}. These studies revealed that even in subsolar-metallicity discs, the formation of giant planets can occur through the disc fragmentation. \cite{Cai2006} found that the subsolar-metallicity disc is gravitationally unstable, although disc fragmentation does not occur. However, such previous studies only followed the evolution for much shorter periods than the lifetime of protoplanetary discs of $\sim3$--$6$\:Myr \citep[e.g.][]{Haisch2001, Hernandez2007, Mamajek2009, Ribas2014}. \citet{Vorobyov2020-9} followed the longest-term evolution among the previous studies, for $\sim$0.4\:Myr elapsed from the disc formation and found that gravitational fragmentation does occur in the range of 0.1--0.01\:$\Zsun$, although this phase is shorter in discs with 0.01\:$\Zsun$ provided that stellar masses are similar.

%---------------------------------------------------%

\cite{Vorobyov2013} performed long-term ($>1$\:Myr) numerical hydrodynamics simulations of the disc evolution in the thin-disc limit, albeit with a metallicity of 1.0\:$\Zsun$. This study identified surviving fragments formed through gravitational instability until the end of simulations in six models. The masses of the surviving fragments were $3.5$--$43$\:$\MJup$, and three fragments were within the planetary mass range ($<13$\:$\MJup$). The fragments with the planetary mass were located between 200 and 400\:au from their host stars. This study concluded that the disc gravitational fragmentation can form a giant planet at distances of $>100$\:au with a metallicity of 1.0\:$\Zsun$, but the formation of a giant planet interior to 100\:au is difficult.

%---------------------------------------------------%

In this work, we perform a two-dimensional thin-disc hydrodynamic numerical simulation to investigate whether giant planets, particularly wide-orbit giant planets, can be formed through the gravitational instability in a protoplanetary disc with a metallicity of $0.1$\:$\Zsun$. Furthermore, we examine whether subsolar protoplanets formed via disc fragmentation can survive for a long-term equivalent to the disc lifetime, and analyze their properties to demonstrate whether the gravitational instability leads to the formation of wide-orbit giant planets. We use a hydrodynamic code similar to that used in \cite{Vorobyov2013}, with some updates related to the thermal and chemical evolution. Especially, we employ a sophisticated thermal scheme, which allows for decoupling of gas and dust temperatures at low metallicities \citep{Vorobyov2020-9, Matsukoba2022}.

%---------------------------------------------------%

This paper is organized as follows. We describe our simulation method and setup in \Secref{Sec:2}. We present the simulation result and explain the evolution and structure of the protoplanetary disc in \Secref{Sec:3}. We then provide the characteristics of a surviving giant planet until the end of our simulation in \Secref{Sec:4}. Discussion and Summary are given in Sections\:\ref{Sec:4} and \ref{Sec:5}, respectively.

%%%%%%%%%%%%%%%%%%%%%%%%%%%%%%%%%%%%%%%%%%%%
%%%%%%%%%%%%%%%%%%%%%%%%%%%%%%%%%%%%%%%%%%%%
%%% SECTION 2 %%%%
\section{Method}
\label{Sec:2}

We perform a two-dimensional simulation integrating the vertical structure of a protoplanetary disc with the metallicity of 0.1\:$\Zsun$ to follow the planet formation via the gravitational instability. Here, we briefly explain the method for the self-gravitational radiation-hydrodynamic simulation and then describe our setup. The details of the method are provided in \cite{Vorobyov2020-6, Vorobyov2020-9}.

%%%%%%%%%%%%%%%%%%%%%%%%%%%%%%%%%%%%%%%%%%%%
%%% SECTION 2-1 %%%%
\subsection{Hydrodynamic simulation}
\label{Sec:2-1}

We use two-dimensional polar-coordinate ($r$, $\phi$) grids with 512$\times$512 spatial zones. The grids are logarithmically spaced in the radial direction and have equal spacing in the azimuthal direction. The computational domain extends to the outer radius of $r_{\rm out}=9900$ \:au. To avoid too small time steps, we set a sink cell at the center of the computational domain with a radius of $r_{\rm sink}=5.2$\:au.

%---------------------------------------------------%

To monitor the time evolution of the surface density, velocity, and internal energy, we solve the vertically integrated mass, momentum, and energy transport equations: 
\begin{align}
&\frac{\partial\Sigma}{\partial t} = - \nabla \cdot \left( \Sigma {\bm u} \right) , 
\label{Eq:mass} \\
&\frac{\partial}{\partial t}\left( \Sigma {\bm u} \right) + \nabla \cdot \left( \Sigma{\bm u} \otimes {\bm u} \right) = -\nabla P + \Sigma{\bm g} + \nabla \cdot {\bm \Pi} , 
\label{Eq:velocity} \\
&\frac{\partial e}{\partial t} + \nabla \cdot \left( e {\bm u} \right) = -P\left( \nabla \cdot {\bm u} \right) - \Lambda + \left( \nabla {\bm u} \right) : {\bm \Pi} ,
\label{Eq:energy}
\end{align}
where $\Sigma$ is the gas surface density, ${\bm u}$ is the planar velocity, $\nabla=\hat{{\bm r}}\partial/\partial r+\hat{{\bm \phi}}r^{-1}\partial/\partial\phi$ is the derivative operator, $P$ is the vertically integrated gas pressure, ${\bm g}$ is the gravitational acceleration, $e$ is the internal energy per unit area, $\Lambda$ is the cooling/heating rate per unit area, and ${\bm \Pi}$ is the viscous stress tensor.
The vertically integrated gas pressure $P$ is calculated by using the ideal-gas equation of state, 
\begin{align}
P = \left( \gamma-1 \right) e ,
\label{Eq:Pressure}
\end{align}
with the adiabatic exponent $\gamma$, which we consistently calculate in accordance with the chemical composition considering the rotational and vibrational degrees of freedom of molecular hydrogen \citep{Omukai1998}. 
The gravitational acceleration ${\bm g}$ consists of two components acting from the gas in the computational domain and the central star in the sink cell. We take into account the turbulent viscosity as a term of the viscous stress tensor ${\bm \Pi}$ by using the $\alpha$ prescription \citep{Shakura1973}. We set the spatially uniform $\alpha$ parameter $\alpha=10^{-3}$. The cooling/heating rate per unit area $\Lambda$ is the sum of the rates of each thermal process \citep{Vorobyov2020-6}. We consider the following thermal processes: the continuum emissions of gas and dust, molecular line emissions of H$_2$ and HD, fine-structure line emissions of O\:{\sc I} ($63\:\mu{\rm m}$) and C\:{\sc II} ($158\:\mu{\rm m}$), and chemical cooling/heating associated with H ionization/recombination and  H$_2$ dissociation/formation. When we calculate the cooling rate of the dust continuum emission, we consider the energy balance on dust grains due to the thermal emission, absorption, and collision with gas, and obtain the dust temperature from this balance. This approach permits decoupling of the gas and dust temperatures in the low-density or high-temperature regime.

%---------------------------------------------------%

We assume that the protostar is located at the center of the sink cell. We initially set the stellar mass to zero, and the protostar grows by accreting the gas flowing into the sink cell at each time step. When the rate of dust continuum emission is calculated, we consider the effect of stellar irradiation which contributes to the thermal stabilization of the disc with the metallicities of $1$--$0.1$\:$\Zsun$ \citep{Matsukoba2022}. In order to include the stellar irradiation, we evaluate the protostellar radius and luminosity composed of two terms, accretion luminosity and stellar intrinsic luminosity. The stellar intrinsic luminosity is calculated from the pre-main-sequence stellar evolutionary track of \cite{DAntona1997}. 

%---------------------------------------------------%

We solve the non-equilibrium chemical network for eight species, H, H$_2$, H$^+$, H$^-$, D, HD, D$^+$, and e$^-$, with 27 reactions (see \citealt{Vorobyov2020-6}). The evolution of the chemical components affects the cooling rates due to molecular line emissions and chemical cooling. We need the chemical fractions of C\:{\sc II} and O\:{\sc I} in order to evaluate the cooling rates of fine-structure line emissions. We assume that all the gas phase C and O are in the form of C\:{\sc II} and O\:{\sc I}, respectively. To perform the simulation for the metallicity of 0.1\:$\Zsun$, we adopt one-tenth of the standard values of the metal contents in the solar neighbourhood for the case of 1\:$\Zsun$: $y_{\rm C}=9.27\times10^{-6}$ and $y_{\rm O}=3.57\times10^{-5}$, where $y_{\rm i}\equiv n_{\rm i}/n_{\rm H}$ is the chemical fraction defined by the ratio of the number densities of a species ${\rm i}$ $n_{\rm i}$ and the hydrogen nuclei $n_{\rm H}$. We also assume neutrality for helium, and the fractional abundance of the helium nuclei is $y_{\rm He}=8.333\times10^{-2}$. The fraction of the deuterium nuclei is $y_{\rm D}=3\times10^{-5}$. Regarding dust grains, we use the standard composition and size distributions in the Galactic interstellar medium \citep{Mathis1977}. We presume the dust-to-gas mass ratio of 0.001, which is ten times smaller than for 1\:$\Zsun$.

%%%%%%%%%%%%%%%%%%%%%%%%%%%%%%%%%%%%%%%%%%%%
%%% SECTION 2-2 %%%%
\subsection{Setup of the simulation}
\label{Sec:2-2}

The numerical simulation starts from the gravitational collapse of a pre-stellar cloud core. The initial surface density and angular velocity profiles of the cloud core we set are derived from an axisymmetric cloud collapse where the angular momentum remains constant \citep{Basu1997}:
\begin{align}
&\Sigma = \frac{\Sigma_0}{\sqrt{1+(r/r_0)^2}} , \\
&\Omega = 2\Omega_0\left( \frac{r_0}{r} \right)^2 \left[ \sqrt{1+\left( \frac{r}{r_0} \right)^2} -1 \right] .
\end{align}
The above profile has a plateau with a uniform surface density extending to the radius $r_0$, and $\Sigma_0$ and $\Omega_0$ are the surface density and angular velocity at the plateau, respectively. We set $\Sigma_0=0.099$\:g\:cm$^{-2}$, $\Omega_0=2.2$\:km\:s$^{-1}$\:pc$^{-1}$, and $r_0=1700$\:au. The mass of the cloud core is 1.0\:$\Msun$ with such a setup. Additionally, the cloud core is designed such that the ratio of the cloud thermal and gravitational energies is 0.78 and the ratio of the rotational and gravitational energy is $7.8\times10^{-3}$.

Since our cloud core is gravitationally unstable, it contracts immediately after starting the simulation. The angular momentum of the infalling gas gradually increases owing to the conservation of angular momentum. The centrifugal radius of the infalling gas becomes larger than the sink cell radius of 5.2\:au after about twice the free-fall time elapsed from the onset of cloud collapse. Subsequently, the protoplanetary disc appears around the sink cell. The size and mass of the disc increase as gas accretes from the envelope, causing the disc to become gravitationally unstable. We observe the evolution of the disc and fragments formed by gravitational instability for 1.0\:Myr after the disc formation, which is a notable fraction of the lifetime of protoplanetary discs of $\sim3$--6\:Myr \citep[e.g.][]{Haisch2001, Hernandez2007, Mamajek2009, Ribas2014}. Such a long-term numerical calculation is computationally expensive. Our current simulation has taken several months on the Intel Xeon Platinum 8174 processors with 48 physical cores, comprising sixty million time steps. We investigate the disc fragmentation and resulting possible planet formation from these results.

%%%%%%%%%%%%%%%%%%%%%%%%%%%%%%%%%%%%%%%%%%%%
%%%%%%%%%%%%%%%%%%%%%%%%%%%%%%%%%%%%%%%%%%%%
%%% SECTION 3 %%%%
\section{Protoplanetary-disc evolution and structure}
\label{Sec:3}

%%%%% FIGURE 1 %%%%%
\begin{figure*}
 \begin{center}
 \begin{tabular}{c} 
  {\includegraphics[width=1.3\columnwidth]{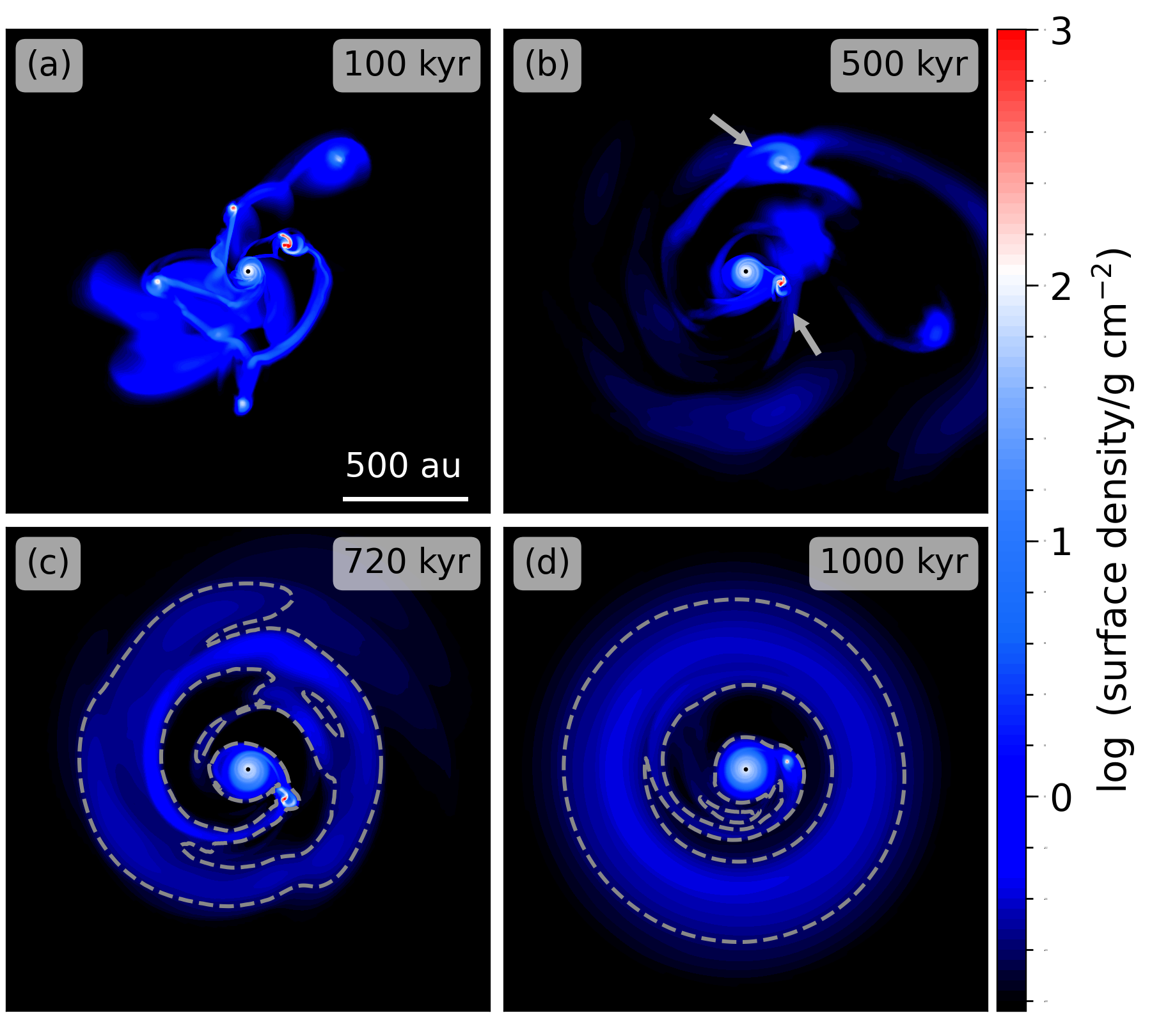}}
 \end{tabular}
 \caption{
 Spatial distribution of the surface density at four different times, (a) 100, (b) 500, (c) 720, and (d) 1000\:kyr after the disc formation. 
 The gray dashed contour lines in panels (c) and (d) represent iso-surface-density lines with 0.25\:g\:cm$^{-2}.$
 The gray arrows in panel (b) indicate the two fragments that are precursors to the surviving protoplanet seen in panels (c) and (d). 
 }
 \label{Fig:1}
 \end{center}
\end{figure*}
%

%---------------------------------------------------%

To illustrate the evolution of the disc, we present the spatial distribution of the surface density at the disc scales at four different epochs in \Figref{Fig:1}. In the early period, several hundred thousand years after the disc formation, the disc fragmentation is vigorous, and multiple fragments can be seen 200--500\:au away from the center. After 500\:kyr from the disc formation, the fragmentation becomes less active, and the number of fragments decreases to a few owing to frequent merging and tidal destruction of the fragments that are gravitationally scattered toward the star. The details of the latter process were investigated in \cite{Vorobyov2020-9}. At 720 and 1000\:kyr, only one fragment orbits around the central star. \Panelsref{Fig:1}{c}{Fig:1}{d} show the same fragment, but at different evolutionary epochs. It is formed by merging of the two fragments indicated by the arrows in \Figref{Fig:1}b. Hereafter, we refer to the fragment as the {\it protoplanet}, and its properties are given in \Secref{Sec:4}.

%---------------------------------------------------%

%%%%% FIGURE 2 %%%%%
\begin{figure}
 \begin{center}
 \begin{tabular}{c} 
  {\includegraphics[width=0.95\columnwidth]{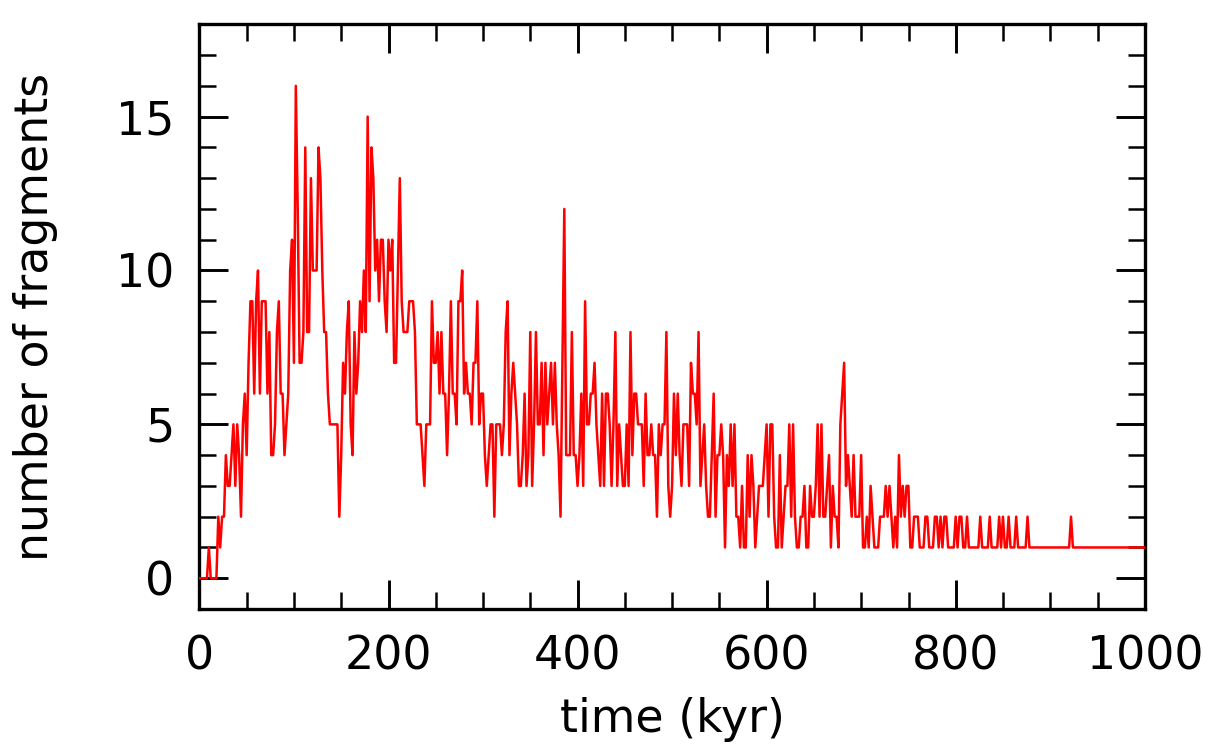}}
 \end{tabular}
 \caption{
 Time evolution of the number of fragments for 1000\:kyr after the disc formation. 
 }
 \label{Fig:2}
 \end{center}
\end{figure}
%

%---------------------------------------------------%

As \Figref{Fig:1} demonstrates, the disc violently fragments in the early phase, and then it gradually stabilizes. To confirm the time variation of the disc fragmentation in more detail, we examine the number of fragments at each snapshot. We use a fragment-tracking algorithm described in \cite{Vorobyov2013}. \Figref{Fig:2} shows the time evolution of the number of fragments. During the first 100\:kyr after the disc formation, the number of fragments increases to a maximum of 16. Until 200\:kyr, the number remains around 10. The strength of gravitational instability decreases with time because the disc mass declines due to accretion on to the central star, and the number of fragments fluctuates around 5 until 550\:kyr, when the protoplanet forms. After 550\:kyr, the number of fragments is usually fewer than 5, and after $\sim$900\:kyr, only one fragment survives. In later time ($\gtrsim750$\:kyr), a second fragment forms episodically. This is formed by the fragmentation of a circumplanetary disc. We explain the density distribution of the circumplanetary disc and its fragmentation in \Secref{Sec:4-2}. 

%---------------------------------------------------%

%%%%% FIGURE 3 %%%%%
\begin{figure*}
 \begin{center}
 \begin{tabular}{c} 
  {\includegraphics[width=1.85\columnwidth]{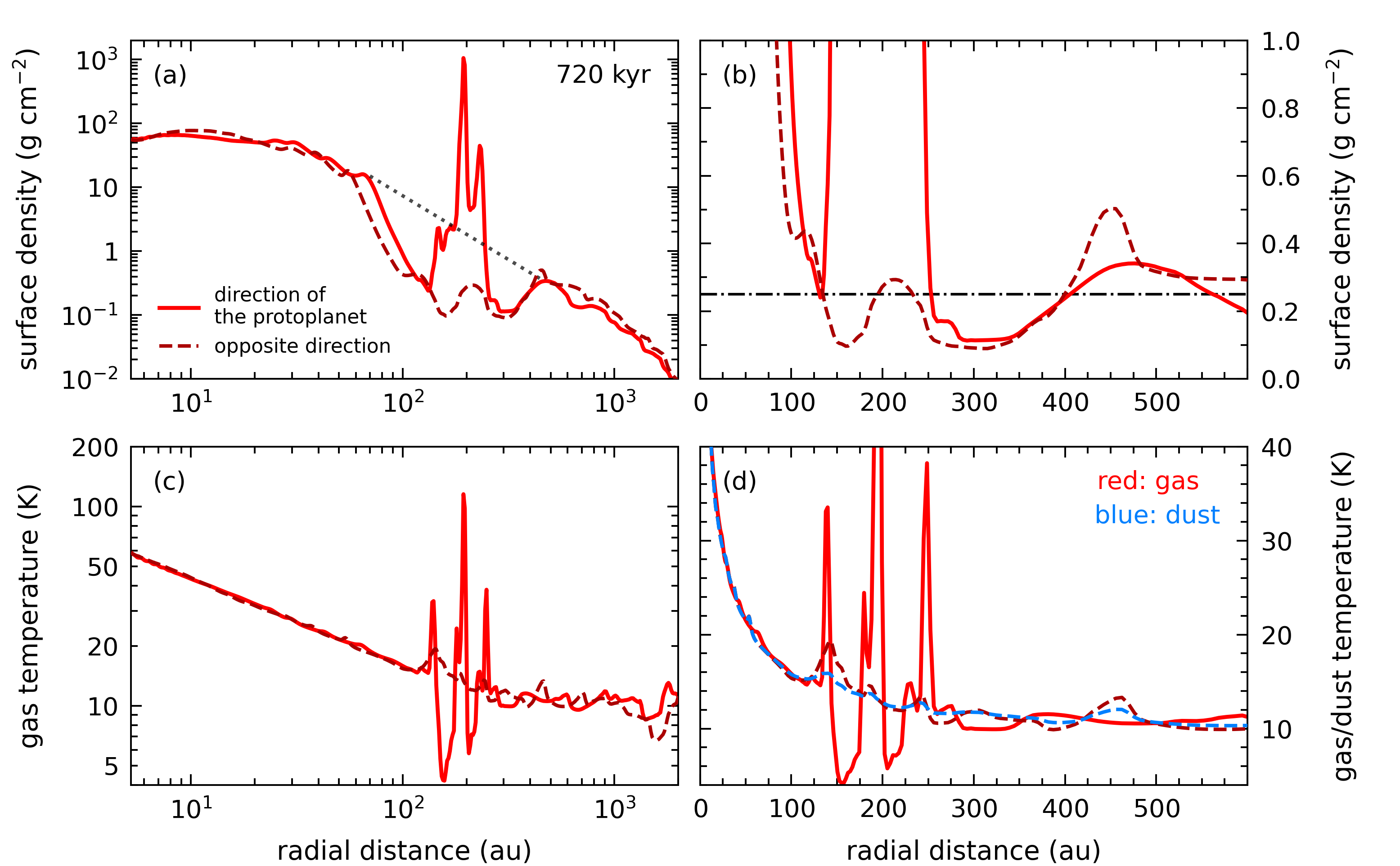}}
 \end{tabular}
 \caption{
 Radial profiles of the surface density (panels a and b) and temperature (panels c and d) across the gap at 720\:kyr elapsed from the disc formation. 
 Panels (a) and (c) are plotted on a logarithmic scale, while panels (b) and (d) are drawn on a linear scale. 
 The solid and dashed lines represent the profiles in the direction of the protoplanet and the opposite direction, respectively. 
 The dotted line in panel (a) is drawn by assuming that the densities inside and outside the gap are connected by a power law, $\Sigma\propto r^{-2}$.
 The dotted-dashed line in panel (b) denotes 0.25\:g\:cm$^{-2}$, as shown by the contours in \Panelsref{Fig:1}{c}{Fig:1}{d}.
 The colours in panel (d) indicate the gas (red) and dust temperatures (blue).
 }
 \label{Fig:3}
 \end{center}
\end{figure*}
%

%---------------------------------------------------%

The gray-dashed lines in \Panelsref{Fig:1}{c}{Fig:1}{d} represent the iso-surface-density contours of 0.25\:g\:cm$^{-2}$. These panels show that there is a density gap in the vicinity of the orbit of the protoplanet. Some previous numerical simulations have reported that a density gap is formed by the orbital motion of an embedded planet in a protoplanetary disc \citep[e.g.][]{Fung2014, Duffell2015, Kanagawa2016}. In \Panelsref{Fig:3}{a}{Fig:3}{b}, we present the radial distributions of the surface density in the direction of the protoplanet (solid line) and in the opposite direction (dashed line) at 720\:kyr corresponding to \Figref{Fig:1}c. There is a density spike at $\sim$200\:au, which corresponds to the protoplanet (solid line in \Figref{Fig:3}a). The density drops both inside and outside the protoplanet's position ($\sim$70 -- 400\:au), and that clearly indicates a density gap. \Figref{Fig:3}b shows the same as \Figref{Fig:3}a, but plotted on a linear scale. The dashed line in \Figref{Fig:3}b indicates that the density is $\sim0.1$\:g\:cm$^{-2}$ within the gap. In order to estimate the depth of the gap, we draw an auxiliary line, which connects the inner and outer border of the gap with a single power law, in \Figref{Fig:3}a. The auxiliary line indicates that the density is $\sim$1\:g\:cm$^{-2}$ around 200\:au. That means the depth of the gap is one order of magnitude of the density.

%---------------------------------------------------%

\Panelsref{Fig:3}{c}{Fig:3}{d} display the radial profiles of temperature. Three temperature spikes are seen between 100 and 300\:au in the protoplanet direction (solid line). The location of the middle spike coincides with the density spike of the protoplanet, and that indicates this spike corresponds to the protoplanet. The positions of the inner and outer spikes match those of the tail of the density spike. The inner and outer tails of the density spike correspond to the edges of the circumplanetary disc. The circumplanetary disc rotates around the protoplanet, and its edges are compressed by the surrounding gas. This compressional heating causes the temperature at the edges to rise, forming the spikes. The linear scale plot in \Figref{Fig:3}d also shows the dust-temperature radial profile in the opposite direction (blue-dashed line). The difference between the gas and dust temperatures is small and its maximum is a few degrees (see dashed lines in \Figref{Fig:3}d). Therefore, the thermal model in previous studies of discs with $\ge$0.1\:$\Zsun$, which assumes the equality between the gas and dust temperatures, serves as a good approximation for understanding the temperature structure of the gap. This, however, may not be true if metallicity decreases below 0.1\:$\Zsun$ \citep{Vorobyov2020-9}.

%%%%%%%%%%%%%%%%%%%%%%%%%%%%%%%%%%%%%%%%%%%%
%%%%%%%%%%%%%%%%%%%%%%%%%%%%%%%%%%%%%%%%%%%%
%%% SECTION 4 %%%%
\section{Surviving Protoplanet}
\label{Sec:4}

Here, we focus on the protoplanet identified in \Secref{Sec:3}. We analyze the evolution of the mass, orbital distance, and radius of the protoplanet to reveal its characteristics.

%%%%%%%%%%%%%%%%%%%%%%%%%%%%%%%%%%%%%%%%%%%%
%%% Section 4.1 %%%%
\subsection{Protoplanet formation and evolution}
\label{Sec:4-1}

%%%%% FIGURE 4 %%%%%
\begin{figure*}
 \begin{center}
 \begin{tabular}{c} 
  {\includegraphics[width=1.3\columnwidth]{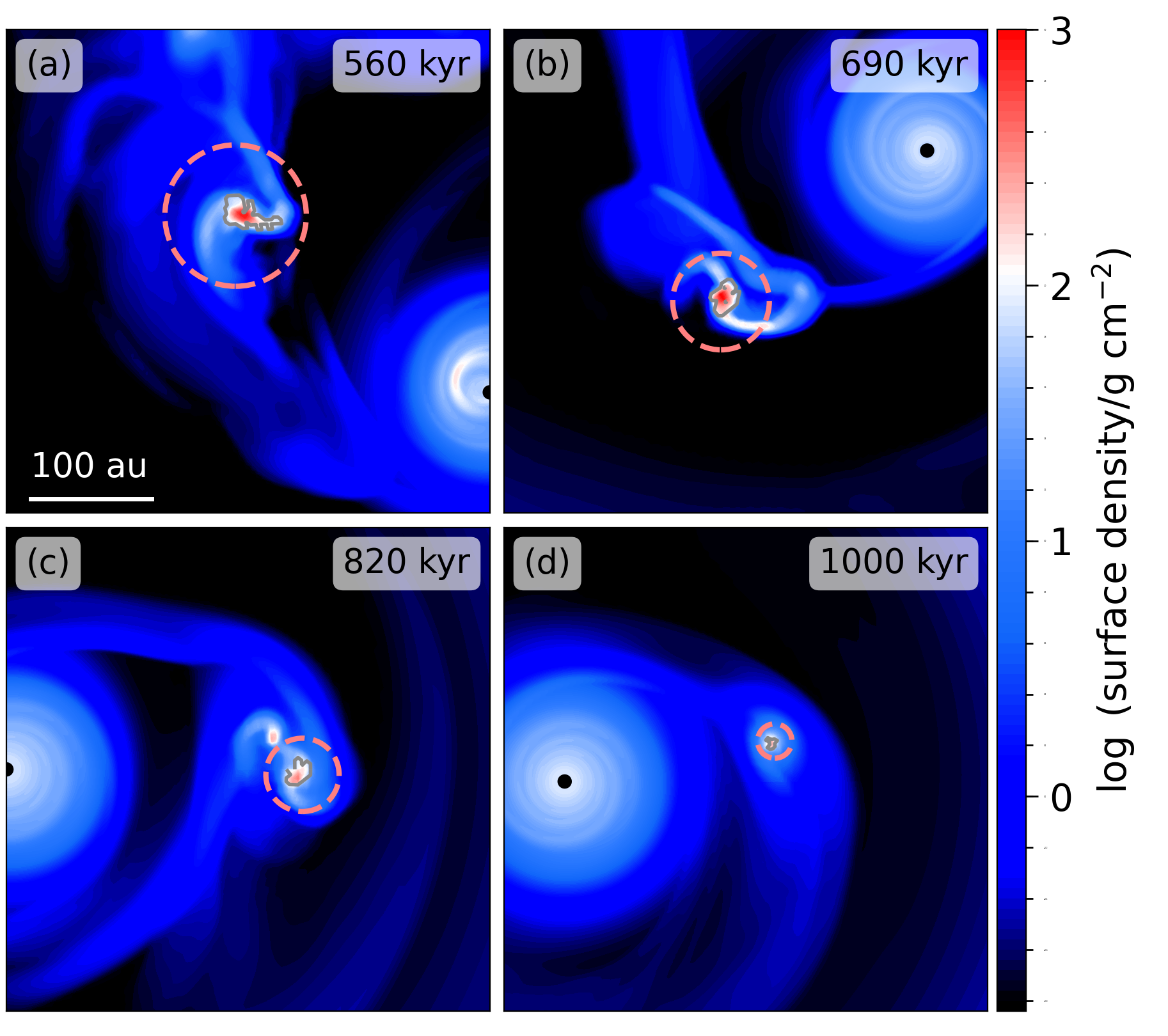}}
 \end{tabular}
 \caption{
 Zoomed-in view on the surviving protoplanet at four different times, (a) 560, (b) 690, (c) 820, and (d) 1000\:kyr after the disc formation. 
 In each panel, the gray curve outlines the protoplanet, and the red-dashed circle shows the Hill radius. 
 }
 \label{Fig:4}
 \end{center}
\end{figure*}
%

%---------------------------------------------------%

\Figref{Fig:4} shows the spatial distribution of the surface density around the protoplanet at four different epochs from immediately after the protoplanet is formed to the end of the calculation. The protoplanet is formed at 550\:kyr by the merger of the two fragments indicated by the arrows in \Figref{Fig:1}b. After that, as shown in \Figref{Fig:4}, the protoplanet survives while orbiting around the central star for 450\:kyr until 1000\:kyr. The gray curves in \Figref{Fig:4} represent the shape of the protoplanet identified by our clump-tracking scheme \citep{Vorobyov2013}. The spatial scale of the protoplanet progressively decreases during this period. The red-dashed circle corresponds to the Hill radius $R_{\rm H}$:
\begin{align}
R_{\rm H} = r_{\rm p}\left( \frac{M_{\rm p}}{3M_{\ast}} \right)^{\frac{1}{3}} ,
\label{Eq:R_Hill}
\end{align}
where $r_{\rm p}$ is the orbital distance of the protoplanet, $M_{\rm p}$ is the protoplanet mass confined inside the gray curves, and $M_{\ast}$ is the central stellar mass, the value of which is stays close to 0.4\:$\Msun$. Therefore, the time variation of the Hill radius depends only on the orbital radius and mass of the protoplanet. \Figref{Fig:4} shows that the Hill radius also gradually shrinks similar to the spatial scale of the protoplanet. The Hill radius describes the scale of a protoplanet's sphere of gravitational influence. This implies that both the scale of the protoplanet and that of the circumplanetary disc influenced by the protoplanet's gravity shrink.

%---------------------------------------------------%

%%%%% FIGURE 5 %%%%%
\begin{figure*}
 \begin{center}
 \begin{tabular}{c} 
  {\includegraphics[width=1.85\columnwidth]{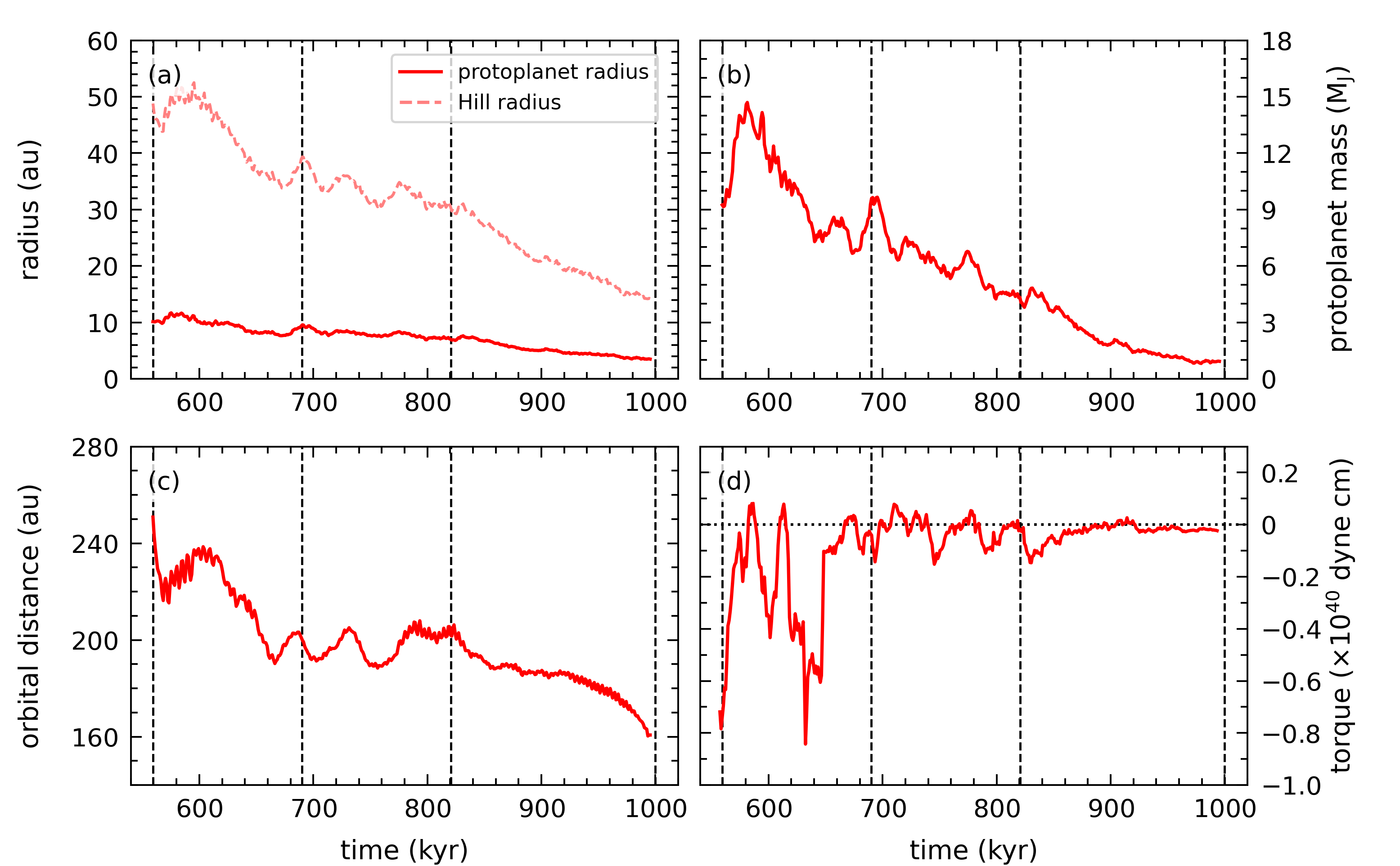}}
 \end{tabular}
 \caption{
 Time evolution of the protoplanet characteristics, (a) the protoplanet radius and Hill radius, (b) protoplanet mass, (c) orbital distance, and (d) gravitational torque acting on the centre of the protoplanet.
 The line represents the time-averaged value using a running average with a time window of 16\:kyr, which corresponds to four Keplerian orbital periods at 200\:au.
 The vertical dashed lines depict the epochs considered in \Figref{Fig:4}.
 The line styles in panel (a) correspond to the protoplanet radius (solid) and the Hill radius (dashed). 
 The horizontal dotted line in panel (d) indicates the zero torque.
 }
 \label{Fig:5}
 \end{center}
\end{figure*}
%

%---------------------------------------------------%

\Figref{Fig:5}a illustrates the time evolution of both the protoplanet radius and Hill radius, where the protoplanet radius is obtained by drawing a circle with an effective area equal to the one enclosed by the gray curves in \Figref{Fig:4}. We often found it challenging to distinguish between the protoplanet and its circumplanetary disc due to their highly dynamic evolution, leading to substantial short-term fluctuations in the derived values. We present only time-averaged values to minimize these fluctuations. We further discuss the accuracy of distinguishing between the protoplanet and the circumplanetary disc for our fragment-tracking algorithm in \Secref{Sec:5-1}. As already seen in \Figref{Fig:4}, both the protoplanet radius and Hill radius decrease over time. Specifically, they reduce from 10 to 4\:au and from 50 to 14\:au, respectively.

%---------------------------------------------------%

We present the time evolution of the protoplanet mass and orbital distance, which are used in estimating the Hill radius, in \Panelsref{Fig:5}{b}{Fig:5}{c}. \Figref{Fig:5}b depicts the protoplanet mass, which is the total mass of gas within the gray curve in \Figref{Fig:4}. The mass is $\sim10$\:$\MJup$ immediately after the formation of the protoplanet and varies between 10--15\:$\MJup$ for 50\:kyr. After 600\:kyr, the mass tends to decline with some temporal variations. It eventually becomes $\sim1$\:$\MJup$ at 1000\:kyr. \Figref{Fig:5}c shows the orbital distance of the protoplanet. The fledgling protoplanet orbits at $\sim250$\:au from the center. After 600\:kyr, the protoplanet migrates inward slowly and reaches 160\:au at 1000\:kyr. 
The mass and orbital distance of the protoplanet decrease to one-tenth and two-thirds of their initial values, respectively. Although the dependence of the Hill radius on the protoplanet mass is weak, the effect of decreasing protoplanetary mass is greater compared to the decrease in the orbital distance, indicating that the primary cause for the shrinking of the Hill radius is the mass loss of the protoplanet.

%---------------------------------------------------%

\Figref{Fig:4} demonstrates that the circumplanetary disc extends beyond the Hill radius. The circumplanetary disc outside the Hill radius is prone to be stripped by the tidal forces of the central star, resulting in a reduction of the mass of the circumplanetary disc. This leads to a decrease in the density of the vicinity of the protoplanet and a subsequent reduction in the external pressure on the protoplanet, thus facilitating mass loss from the outer layers of the protoplanet to its surroundings.

%---------------------------------------------------%

The protoplanet slowly moves inward from 250 to 160\:au for $\sim$450\:kyr as shown in \Figref{Fig:5}c. To understand this orbital evolution, we analyze the gravitational torques. The gravitational torque acting on the protoplanet $\mathcal{T}$ is written as
\begin{align}
\mathcal{T} = \sum_{\rm i} r_{\rm p}|{\bm F}_{\rm i}| \sin\gamma_{\rm i} ,
\label{Eq:Torque}
\end{align}
where $|{\bm F}_{\rm i}|$ is the absolute value of the gravitational force acting on the protoplanet from the ${\rm i}$-th grid cell, and $\gamma_{\rm i}$ is the angle between the position vector of the protoplanet ${\bm r}_{\rm p}$ and the gravitational force ${\bm F}_{\rm i}$. 
The absolute value of the gravitational force from the ${\rm i}$-th grid cell is given by
\begin{align}
|{\bm F}_{\rm i}| = G\frac{M_{\rm p}m_{\rm i}}{R_{\rm i}^2} ,
\label{Eq:GForce}
\end{align}
where $G$ is the gravitational constant, $m_{\rm i}$ is the gas mass located in the ${\rm i}$-th grid cell, and $R_{\rm i}$ is the distance between the protoplanet and the ${\rm i}$-th grid cell. We ignore the gravitational torque from the grid cells within twice the Hill radius because we focus on the effect of the gravity from the protoplanetary-disc scale. The time evolution of the gravitational torque is shown in \Figref{Fig:5}d. The value of the torque is negative most of the time. That is consistent with the inward migration of the protoplanet, and we argue that the migration is induced by the gravitational torque from the protoplanetary disc. In the earlier stage, the amplitude of the torque is larger due to the stronger gravitational force from the higher protoplanet mass. 

%---------------------------------------------------%

We can estimate the migration timescale of the protoplanet by using the gravitational torque obtained from \Eqref{Eq:Torque}. The migration timescale $t_{\rm mg}$ is calculated using the simple equation \citep{Vorobyov2013}
\begin{align}
t_{\rm mg} = \frac{L}{-2\mathcal{T}} ,
\label{Eq:migration}
\end{align}
where $L=r_{\rm p}M_{\rm p}v_{\rm p}$ is the angular momentum of the protoplanet and $v_{\rm p}$ is the orbital velocity of the protoplanet. We use the Keplerian velocity as the orbital velocity for simplicity. For the orbital distance $r_{\rm p}$, protoplanet mass $M_{\rm p}$, and gravitational torque $\mathcal{T}$, we use the time-averaged values because they highly fluctuate over time. For instance, between 690 and 820\:kyr (from \Figref{Fig:4}b to \Figref{Fig:4}c), the time-averaged values of the angular momentum and gravitational torque are $4.7\times10^{51}$\:g\:cm$^2$\:s$^{-1}$ and $-1.6\times10^{38}$\:dyne\:cm, respectively. Consequently, from \Eqref{Eq:migration}, the migration timescale is 480\:kyr. This value is comparable to the evolutionary time period of the protoplanet after its birth, 450\:kyr. Its duration is consistent with the fact that the protoplanet gradually migrates inward for 450\:kyr.

%%%%%%%%%%%%%%%%%%%%%%%%%%%%%%%%%%%%%%%%%%%%
%%% Section 4.2 %%%%
\subsection{Circumplanetary disc}
\label{Sec:4-2}

%---------------------------------------------------%

Recent observational developments have improved our understanding of the environment of wide-orbit giant planets ($\gtrsim10$\:au). Several circumplanetary discs have been observed, including some candidates. For instance, the PDS\:70 system is composed of a young T Tauri star and two planets, PDS\:70\:b and c \citep{Keppler2018, Muller2018, Haffert2019}. H$\alpha$ line emission has been detected from both planets, indicating that gas accretion occurs on the planets, which is an indirect evidence for a circumplanetary disc \citep{Wagner2018, Haffert2019}. In addition, a compact ($<1.2$\:au) dust continuum emission has been found in PDS\:70\:c \citep{Isella2019, Benisty2021}. This provides a direct image of a circumplanetary disc. Other candidates for circumplanetary discs have been also reported so far \citep{Mohanty2007, Quanz2015, Bae2022}.

%---------------------------------------------------%

%%%%% FIGURE 6 %%%%%
\begin{figure*}
 \begin{center}
 \begin{tabular}{c} 
  {\includegraphics[width=1.3\columnwidth]{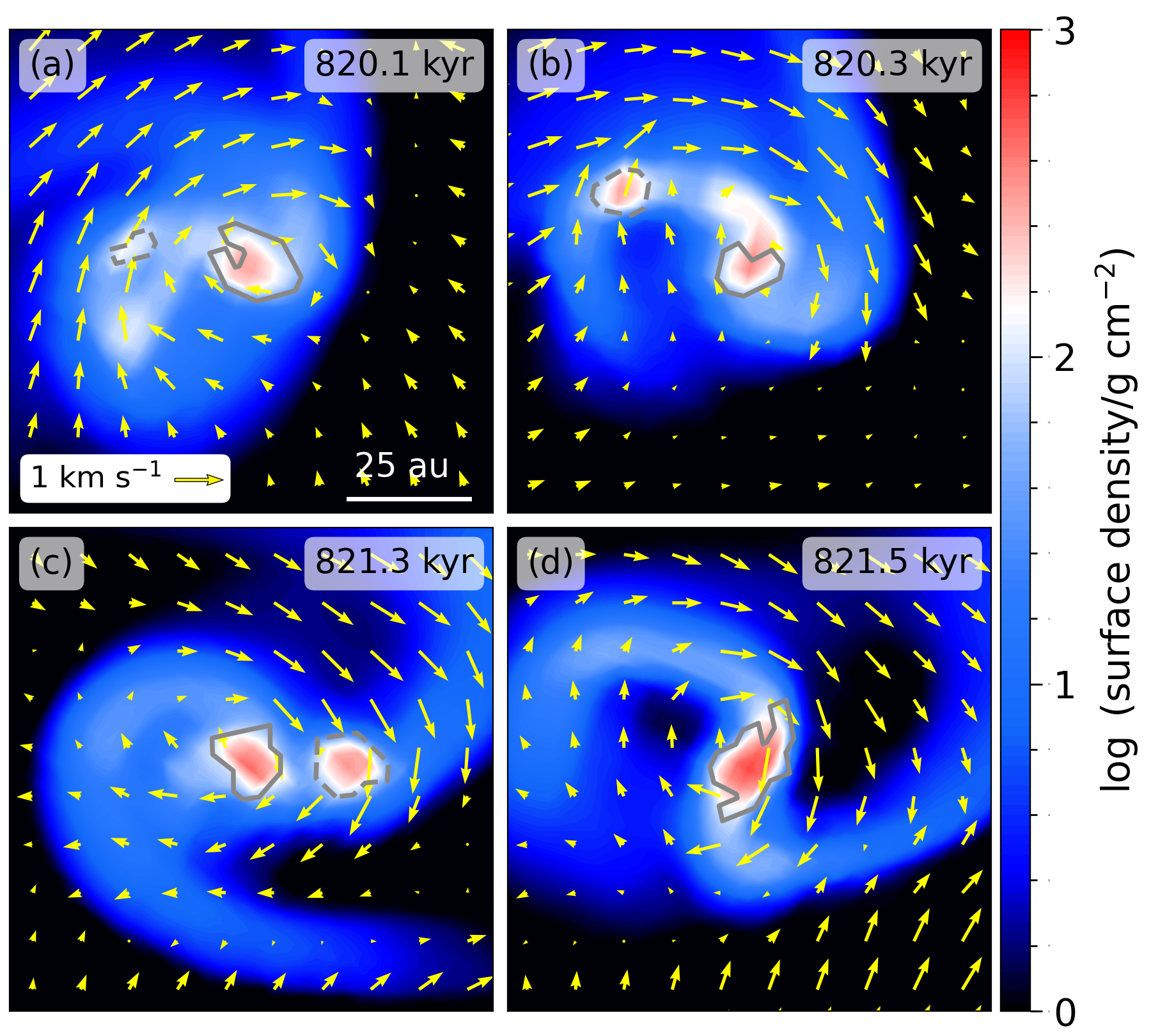}}
 \end{tabular}
 \caption{
 Snapshots of the vicinity of the protoplanet from the moment of the circumplanetary disc fragmentation to that of the merger of the protoplanet and a fragment. 
 The gray-solid and dashed curves delineate the protoplanet and the fragment in the circumplanetary disc, respectively. 
 The yellow arrows represent the relative velocity to the density peak of the protoplanet.
 }
 \label{Fig:6}
 \end{center}
\end{figure*}
%

%---------------------------------------------------%

We find a substructure with a surface density of $\gtrsim$10\:g\:cm$^{-2}$ around the protoplanet in all the snapshots shown in \Figref{Fig:4}. The scale of the substructure is similar or greater than the Hill radius, and this indicates that at least part of the substructure is  gravitationally bound to the protoplanet. \Figref{Fig:6} displays the spatial distribution of surface density near the protoplanet on a smaller spatial scale than shown in \Figref{Fig:4}, and we confirm that the substructure is not just gas around the protoplanet but a circumplanetary disc from \Figref{Fig:6}. The shown epoch in \Figref{Fig:6}a is immediately after that in \Figref{Fig:4}c. The yellow arrows represent the gas velocity field in the frame of reference of the density peak of the protoplanet. In \Figref{Fig:6}a, the gas around the protoplanet rotates clockwise with respect to the protoplanet (the counter direction to the orbiting motion of the protoplanet), meaning that the surrounding material is gravitationally bound to the fragment. Thus, we regard the gas around the protoplanet as the circumplanetary disc. 

%---------------------------------------------------%

The shape of the circumplanetary disc is not a neat circle around the protoplanet, but rather a non-axisymmetric structure (\Figref{Fig:6}a). In particular, we see a spiral arm-like structure from the left to the bottom left of the protoplanet in \Figref{Fig:6}a. There is also a region with a local density enhancement within this spiral arm (gray-dashed line in \Figref{Fig:6}a). The local region increases in density with time and causes gravitational fragmentation of the circumplanetary disc (\Panelsref{Fig:6}{b}{Fig:6}{c}). Eventually, the fragment  merges with the protoplanet after circling the protoplanet (\Figref{Fig:6}d). 

%---------------------------------------------------%

%%%%% FIGURE 7 %%%%%
\begin{figure}
 \begin{center}
 \begin{tabular}{c} 
  {\includegraphics[width=0.95\columnwidth]{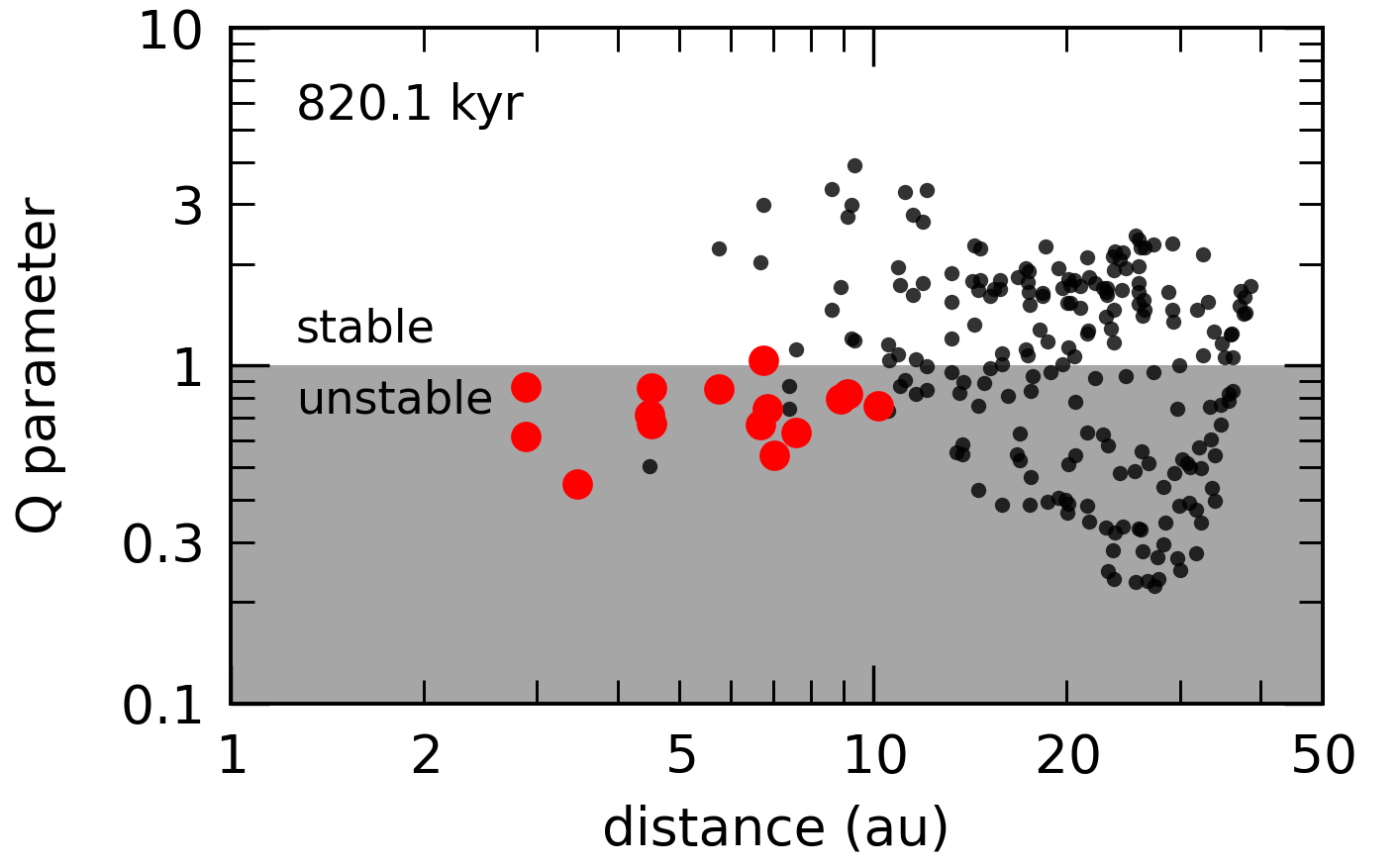}}
 \end{tabular}
 \caption{
 Toomre's Q parameter distribution in the circumplanetary disc at 820.1\:kyr, the same epoch as in \Figref{Fig:6}a. 
 The horizontal axis is the distance from the density-peak of the protoplanet.
 The red points represent the grid cells within the protoplanet, and the black points express those with the surface density of $\ge10$\:g\:cm$^{-2}$.
 The gray shaded region indicates that the Q parameter is less than unity, signifying that the circumplanetary disc is gravitationally unstable and prone to fragmentation.
 }
 \label{Fig:7}
 \end{center}
\end{figure}
%

%---------------------------------------------------%

We evaluate the Toomre's Q parameter \citep{Toomre1964} of the circumplanetary disc in order to confirm that the fragmentation is caused by gravitational instability. The Q parameter $\mathcal{Q}_{\rm T}$ is given by, 
\begin{align}
\mathcal{Q}_{\rm T} = \frac{c_{\rm s}\Omega'}{\pi G\Sigma} ,
\end{align}
where $c_{\rm s}$ is the sound speed and $\Omega'$ is the angular velocity in the frame of reference of the density peak of the protoplanet. The sound speed is defined by $c_{\rm s}=\sqrt{\gamma P/\Sigma}$, and the angular velocity is written by $\Omega'=u'_{\rm rot}/R'$, where $u'_{\rm rot}$ is the rotating velocity in the frame of reference of the density peak of the protoplanet and $R'$ is the distance from the density peak of the protoplanet. 
The Q parameters for the grid cells with the surface density of $\ge10$\:g\:cm$^{-2}$ around the protoplanet are shown in \Figref{Fig:7}. Some black points, which have values less than unity and are distributed around 20\:au, correspond to the spiral arm-like structure seen in \Figref{Fig:6}a, meaning that this structure is gravitationally unstable, and the circumplanetary disc fragments by gravitational instability.

%---------------------------------------------------%

The fragmentation of the circumplanetary disc frequently occurs at other epochs. The formation of a fragment and its disappearance by merging with the protoplanet is a recurrent phenomenon, which explains the appearance of a second planetary object seen after $\sim$750\:kyr in \Figref{Fig:2}.

%---------------------------------------------------%

The mass of the fragment seen in \Figref{Fig:6} is $\sim1\:\MJup$, which is of the same order as the protoplanet with a mass of $\sim4\:\MJup$. Although fragments of the circumplanetary disc do not survive in our calculation, likely because of insufficient numerical resolution to properly follow the protoplanet contraction (see \Secref{Sec:5-1} below), we speculate that another planet similar in mass to the primary could be formed via circumplanetary disc fragmentation. This process may lead to the formation of Jupiter-mass binary planets.

%%%%%%%%%%%%%%%%%%%%%%%%%%%%%%%%%%%%%%%%%%%%
%%%%%%%%%%%%%%%%%%%%%%%%%%%%%%%%%%%%%%%%%%%%
%%% SECTION 5 %%%%
\section{Discussion}
\label{Sec:5}

%%%%%%%%%%%%%%%%%%%%%%%%%%%%%%%%%%%%%%%%%%%%
%%% Section 5.1 %%%%
\subsection{Effect of spatial resolution on mass loss}
\label{Sec:5-1}

%---------------------------------------------------%

A fragment's radius shortly after formation by gravitational instability is much larger than the Jovian radius, as observed in our calculation (see \Figref{Fig:5}a). The newly formed fragment via gravitational instability shrinks in three stages \citep{Decampli1979}. First, the fragment contracts quasi-statically, with the central temperature rising from $\sim100$\:K. When the central temperature of the fragment reaches $\sim2000$\:K, dissociation of hydrogen molecules begins, and the fragment collapses dynamically (the second stage). After the second stage, the fragment becomes compact, and its radius decreases to several times the Jovian radius. In the third stage, the fragment contracts slowly ($10^9$\:yr) and becomes sufficiently compact and less affected by the tidal effect.

%---------------------------------------------------%

The timescale for quasi-static contraction in the first stage depends on the planet mass, composition, and distance from the host star, in the range of $10^3$--$10^6$\:yr. According to \cite{Helled2011}, the larger the mass and the lower the metallicity, the shorter the timescale. In their study, an example with the largest planet mass (7\:$\MJup$) and the lowest metallicity (1/3\:$\Zsun$) contracts to sub-au scale in $\sim3000$\:yr. However, in our calculation, even after $\sim10^3$\:yr since the formation of the protoplanet with a radius of $\sim$10\:au, the time-averaged protoplanet radius does not decrease (\Figref{Fig:5}a). Therefore, it can be seen from our calculation that quasi-static contraction does not occur on such short timescales. Note that \cite{Helled2011} considered an isolated planet, but the protoplanet in our simulation constantly interacts with the surrounding gas. The differences in the situations may affect the presence of the quasi-static contraction.

%%%%% FIGURE 8 %%%%%
\begin{figure}
 \begin{center}
 \begin{tabular}{c} 
  {\includegraphics[width=0.95\columnwidth]{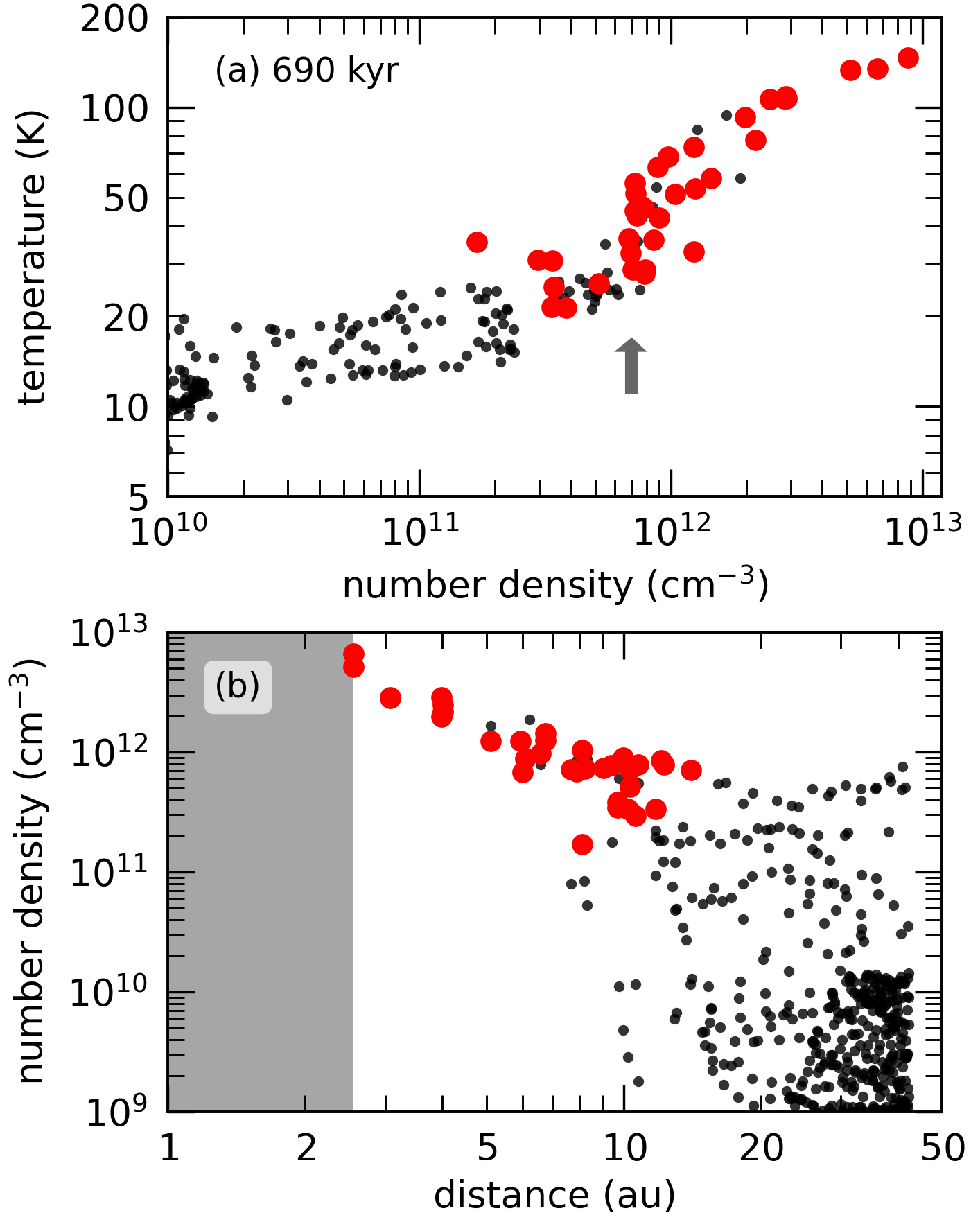}}
 \end{tabular}
 \caption{
 Thermal and density states of the protoplanet and its surrounding at 690\:kyr after the disc formation corresponding to \Figref{Fig:4}b. 
 The upper panel is the temperature-density phase diagram, and the lower panel is the density radial profile from the density-peak grid of the protoplanet.
 The red points represent the grid cells within the protoplanet, and the black points depict those within the Hill radius of 42\:au. 
 The arrow in the upper panel indicates the temperature turning point at which the evolution changes from the isothermal state to adiabatic one. 
 The shaded region indicates our resolution limit, i.e. the distance between the density-peak grid and the adjacent grids is 2.6\:au. 
 }
 \label{Fig:8}
 \end{center}
\end{figure}
%

%---------------------------------------------------%

In \Figref{Fig:8}, we depict the thermal and density states within the Hill radius of the protoplanet at 690\:kyr to clarify why quasi-static contraction does not occur in our calculation. \Figref{Fig:8}a displays the temperature-density phase diagram for the grid cells within the protoplanet (red) and within the Hill radius (black). The temperature within the Hill radius is nearly isothermal, distributed at $10$--$20$\:K for densities of $\lesssim7\times10^{11}$\:cm$^{-3}$. On the other hand, the temperature inside the protoplanet adiabatically increases beyond the density indicated by the gray arrow. The temperature clearly rises from 20\:K to 200\:K for densities of $\gtrsim7\times10^{11}$\:cm$^{-3}$. The differing temperature distributions of the protoplanet and the circumplanetary disc on the phase diagram indicate that we have correctly identified fragments.

%---------------------------------------------------%

\Figref{Fig:8}b shows the radial profile of the density from the density-peak grid of the protoplanet. The grid points that make up the protoplanet exist at radii between 2.6 and 10\:au. There are no grid points within 2.6\:au in the gray-shaded area because this is the minimum radial size of the grids at the position of the protoplanet. We recall that we employ the log-scaled grid in the radial direction, which results in deteriorating numerical resolution with distance from the star. According to \cite{Helled2011}, the protoplanet radius at the end of the quasi-static contraction is sub-au. Our calculation, however, does not have a spatial resolution at sub-au-scales for the protoplanet and cannot track the quasi-static contraction. 

%---------------------------------------------------%

In our simulation, therefore, the protoplanet apparently remains large without undergoing the quasi-static contraction. As a result, the protoplanet orbits while maintaining a larger size than it should have originally, making the protoplanet susceptible to mass loss by the tidal effect. Therefore, the mass of the protoplanet in our simulation should be considered as the minimum value for a giant planet formed by the gravitational fragmentation in a subsolar-metallicity disc. The giant planet without extra mass loss can be more massive up to $\sim$10\:$\MJup$.

%%%%%%%%%%%%%%%%%%%%%%%%%%%%%%%%%%%%%%%%%%%%
%%% Section 5.2 %%%%
\subsection{Comparison with previous numerical works and observations}
\label{Sec:5-2}

%---------------------------------------------------%

Previous numerical studies \citep{Boss2002, Meru2010, Vorobyov2020-9} reported the formation of gaseous clumps via gravitational fragmentation in a subsolar-metallicity protoplanetary disc. The mass of the clumps found in their simulations ranges from $\sim$1 to 20\:$\MJup$, yielding clumps heavier than the protoplanet in our simulation. However, the periods of disc evolution in their simulations, ranging from $\sim$0.5 to 400\:kyr, are shorter than in our simulation, leaving it unclear whether those clumps would survive and, if they do, what their masses would be. Therefore, our study should be considered the first to demonstrate the potential for forming giant planets with masses of $\ge1.0$\:$\MJup$ through gravitational fragmentation in a subsolar-metallicity disc.

%---------------------------------------------------%

The formation of wide-orbit giant planets due to gravitational instability has been investigated through numerical simulations in the case of solar metallicity \citep[e.g.][]{Boss1998, Mayer2007, Zhu2012, Vorobyov2013, Stamatellos2018}. The results of these numerical simulations suggest that gravitational instability is a viable formation scenario. In particular, \cite{Vorobyov2013} conducted numerous numerical calculations with long time periods ($>1$\:Myr) for each model. In this study, the disc evolution was calculated using sixty initial cloud models, and surviving giant planets and brown dwarfs were found in six models. In models where planet formation failed, the fragments either fall into the central star, are ejected from the planetary system, or disappear due to the tidal effect. The six surviving giant planets and brown dwarfs have masses of 3.5--43\:$\MJup$ and orbital distances of 178--415\:au. The results in \cite{Vorobyov2013} indicate that the rate of wide-orbit giant planet formation in a disc with solar metallicity is one in ten. Since our study demonstrated only one model, the formation rate in a subsolar-metallicity disc is unclear. The rate can be obtained by performing numerical calculations with a number of models comparable to those of \cite{Vorobyov2013}. By comparing the formation rate in solar metallicity and subsolar metallicity, we would provide implications for the observational frequency of wide-orbit giant planets in a subsolar-metallicity disc.

%%%%% TABLE 1 %%%%%
\begin{table} 
 \begin{center}
 \caption{Characteristics of wide-orbit giant planets observed by direct imaging with the metallicity of the host star of $[{\rm Fe}/{\rm H}]<-0.3$}
 \label{Tab:1}
  \scalebox{0.97}[0.97]{ 
  {\begin{tabular}{l c c c c} 
     \hline \hline
      & $[{\rm Fe}/{\rm H}]$ & $M_{\ast}$ & $M_{\rm p}$ & $r_{\rm p}$  \\ %\hline
      & & $\Msun$ & $\MJup$ & au  \\ \hline
     kap\:And\:b     & -0.36$^{+0.09}_{-0.09}$ & 2.8$^{+0.1}_{-0.2}$ & 13.0$^{+12.0}_{-2.0}$   & 100.0$^{+27.0}_{-46.0}$  \\     
     HIP\:78530\:b & -0.50$^{+0.03}_{-0.01}$ & 2.5                           & 28.0$^{+10.0}_{-10.0}$ & 710$^{+60}_{-60}$           \\     
     CT\:Cha\:b      & -0.56$^{+0.01}_{-0.01}$ & --                             & 17.0$^{+6.0}_{-6.0}$     & 440.0                                \\    
     HR\:8799\:b    & -0.65$^{+0.02}_{-0.01}$ & 1.56                         & 7.0$^{+4.0}_{-2.0}$       & 68.0                                  \\    
     HR\:8799\:c    & -0.65$^{+0.02}_{-0.01}$ & 1.56                         & 8.3$^{+0.6}_{-0.6}$       & 42.9                                  \\         
     HR\:8799\:d    & -0.65$^{+0.02}_{-0.01}$ & 1.56                         & 8.3$^{+0.6}_{-0.6}$       & 27.0                                  \\       
     HR\:8799\:e    & -0.65$^{+0.02}_{-0.01}$ & 1.56                         & 9.6$^{+1.9}_{-1.9}$       & 16.4$^{+2.1}_{-1.1}$        \\ \hline
  \end{tabular}}
  }
 \end{center}
\end{table}
%

%---------------------------------------------------%

\Tabref{Tab:1} provides the characteristics of the wide-orbit giant planets observed by direct imaging for comparison with those of the planet in our simulation. It lists planets that are found around stars with metallicity of $[{\rm Fe}/{\rm H}]<-0.3$. The metallicity of the host star is taken from \cite{Swastik2021}, and the mass of the host star and the separation between the star and planet are obtained from the Extrasolar Planets Encyclopaedia \footnote{\url{http://exoplanet.eu}}. \Tabref{Tab:1} shows that the lowest mass of the direct-imaged planets is $7$\:$\MJup$ and larger than the protoplanet in our simulation. As discussed in \Secref{Sec:5-1}, suppressing mass loss through the contraction of protoplanets is required in order to form planets with masses of $\ge7$\:$\MJup$. Here, we note that the observational results summarized in \Tabref{Tab:1} do not rule out the existence of planets with masses below 7\:$\MJup$ around subsolar-metallicity stars. Generally, subsolar-metallicity stars are located farther away from the Sun, which introduces a bias making it more challenging to observe direct-imaged planets with lower masses. In addition, the masses of the host stars for the planets listed in \Tabref{Tab:1} are heavier than the mass of the central star in our simulation. Since the mass of a protoplanetary disc tends to increase with the mass of the host star, a planet formed through disc fragmentation in a more massive system might also tend to be heavier.

%---------------------------------------------------%

The direct-imaged planets are located about $15$--$710$\:au away from the host star. The protoplanet in our simulation is located within this range, which is consistent with the observation results. \Tabref{Tab:1} shows that five planets are located within $100$\:au from the host star, but these planets are not reproduced in our simulation or \cite{Vorobyov2013}. One of the scenarios to explain the formation of giant planets with an orbital distance of $\sim$10\:au is the tidal downsizing theory \citep{Nayakshin2010}. In this theory, a clump formed on a 100\:au scale migrates inward. When it reaches $\sim$10\:au, a deep gas gap is opened, slowing the inward migration and causing the clump to remain at that orbital distance. Furthermore, the clump that moves to around 10\:au is stripped of its outer gas by the tidal effect, leaving only the core. In addition, \cite{Vorobyov2018} found that even without gap formation, the tidal mass loss from a clump slows down or halts inward migration of the clump, causing the clump to remain at a radial distance of a few tens of au. The calculations in \cite{Vorobyov2018} have a higher spatial resolution than ours, and therefore a dense core survives after being gas stripped away. Such a dense core would be formed also in a subsolar-metallicity disc by calculations with higher resolution.

%---------------------------------------------------%

HR\:8799 is a planetary system with multiple wide-orbit giant planets. However, in both our study and \cite{Vorobyov2013}, only one surviving planet exists per host star, and a planetary system of multiple giant planets is not formed. In our simulation, two fragments orbit for 100\:kyr during the period between 450 and 550\:kyr before merging. As discussed \Secref{Sec:5-1}, in high-resolution simulations, the size of the fragments may be compact due to the quasi-static contraction. Such compact fragments may be able to avoid merging, hence a system with multiple giant planets could potentially be formed.

%%%%%%%%%%%%%%%%%%%%%%%%%%%%%%%%%%%%%%%%%%%%
%%%%%%%%%%%%%%%%%%%%%%%%%%%%%%%%%%%%%%%%%%%%
%%% SECTION 6 %%%%
\section{Summary}
\label{Sec:6}

We have investigated the formation of a wide-orbit giant planet by gravitational instability in the subsolar-metallicity protoplanetary disc by performing a two-dimensional radiation-hydrodynamic simulation with the metallicity of 0.1\:$\Zsun$. Our simulation follows the long-term disc evolution for 1\:Myr elapsed from disc formation, corresponding to the disc lifetime. We have found a surviving planet and analyzed its properties. Our findings are summarized as follows.

\begin{itemize}
\item[(i)] The protoplanetary disc is gravitationally unstable and violently fragments until 200\:kyr after its formation. Two fragments merge at 550\:kyr, leaving only one fragment that remains in the disc and survives until 1000\:kyr.

\item[(ii)] The surviving protoplanet has a mass of $\sim$10\:$\MJup$ at its birth, and then the mass decreases to 1\:$\MJup$ at 1\:Myr. The protoplanet and circumplanetary disc lose their masses by the tidal effect of the central star. The radius of the protoplanet decreases in concert with the mass loss. 

\item[(iii)] The protoplanet gradually migrates inward from 250 to 160\:au. This radial motion is induced by the gravitational torque on the protoplanetary-disc scale. The orbital distance at the end of the calculation is within the range of separations between host stars and wide-orbit giant planets observed by direct imaging.

\item[(iv)] The protoplanet is surrounded by a circumplanetary disc, which is gravitationally unstable and frequently fragments. The mass of fragments of the circumplanetary disc is $\sim1$\:$\MJup$ and similar to that of the protoplanet. In our simulation, fragments of the circumplanetary disc do not survive because they merge with the protoplanet.
\end{itemize}

Our simulation has demonstrated that a protoplanet with $\ge1$\:$\MJup$ can be formed via gravitational instability in a protoplanetary disc with the metallicity of 0.1\:$\Zsun$ and can survive for a long term until 1\:Myr after the disc formation. Gravitational instability facilitates the formation of giant planets not only in a disc with solar metallicity but also with subsolar metallicity. Our results provide the minimum mass of a giant planet formed through gravitational instability, and we predict that the mass of a giant planet, with mitigated excessive mass loss, could increase to $\sim$10\:$\MJup$. This mass lies within the range of the masses of subsolar-metallicity giant planets observed by direct imaging so far. Our prediction will be confirmed through high-resolution and long-time-span numerical calculations in future studies.

%%%%%%%%%%%%%%%%%%%%%%%%%%%%%%%%%%%%%%%%%%%%%%%%%
%%%%%%%%%%%%%%%%%% ACKNOWLEDGMENTS %%%%%%%%%%%%%%%%%%

\section*{Acknowledgments}

The authors express their cordial gratitude to Takahiro Tanaka for his continuous interest and encouragement. The authors would also like to thank Hidekazu Tanaka and Sanemichi Takahashi for fruitful discussions and useful comments. R.M. and T.H. are grateful for the support by Grants-in-Aid for Scientific Research (19H01934, 21H00041) from the Japan Society for the Promotion of Science. E.I.V. and M.G. acknowledge support by the Austrian Science Fund (FWF) under research grant P31635-N27. The simulations were performed on the Vienna Scientific Cluster (VSC-4).

%%%%%%%%%%%%%%%%%%%%%%%%%%%%%%%%%%%%%%%%%%%%%%%%%
%%%%%%%%%%%%%%%%%%% DATA AVAILABILITY% %%%%%%%%%%%%%%%%%%

\section*{Data availability}

The data underlying this article will be shared on reasonable request to the corresponding author.


%\bibliographystyle{mnras}
%\bibliography{} % if your bibtex file is called example.bib

%%%%%%%%%%%%%%%%%%%%%%%%%%%%%%%%%%%%%%%%%%%%%%%%%%
%%%%%%%%%%%%%%%%%%%%%%%%%%%%%%%%%%%%%%%%%%%%%%%%%%

%%%%%%%%%%%%%%%%%%%%%%%%%%%%%%%%%%%%%%%%%%%%%%
%%%%%%%%%%%%%%%%% APPENDICES %%%%%%%%%%%%%%%%%%%%%
%\appendix

%%%%%%%%%%%%%%%%%%%%%%%%%%%%%%%%%%%%%%%%%%%%%%%%%%

% Don't change these lines
\bsp	% typesetting comment
\label{lastpage}
\end{document}